\documentclass[epj]{svjour}

\usepackage{epsf,latexsym}
\usepackage{graphics}
\usepackage{color}

\newcommand{\be}{\begin{equation}}
\newcommand{\ee}{\end{equation}}
\newcommand{\ba}{\begin{array}{l}}
\newcommand{\ea}{\end{array}}
\newcommand{\re}[1]{(\ref{#1})}

\begin{document}
\title{Particle Transport in Graphene Nanoribbon Driven by Ultrashort Pulses}
\author{D. Babajanov \inst{1}, 
D.U. Matrasulov \inst{1} \and R. Egger \inst{2}}

\offprints{d.matrasulov@polito.uz}      
\institute{Turin Polytechnic University in Tashkent, 17. Niyazov
Str., 100095, Tashkent, Uzbekistan \and Institut f\"{u}r
Theoretische Physik, Heinrich-Heine-Universit\"{a}t, D-40225
D\"{u}sseldorf, Germany}

\abstract{We study charge transport in a graphene zigzag
nanoribbon driven  by an external time-periodic kicking potential. 
Using the exact solution of the time-dependent Dirac equation with a
delta-kick potential acting in each period, we study
the time evolution of the quasienergy levels and the
time-dependent optical conductivity.  By variation of the 
kicking parameters, the conductivity becomes widely tunable.
} 

\maketitle

\section{Introduction}
\label{intro}

Ever since its experimental discovery a decade ago, the physics of graphene has been a
hot topic in condensed matter physics, see 
Refs.~\cite{review1,review2,review3,review4,review5,review6,review7} for reviews. 
One of the most intensely studied class of problems concerns electronic transport in bulk or 
confined graphene monolayers. We here focus on the particle dynamics in externally driven 
graphene samples, where a general goal is to achieve tunability of charge transport.
A rich variety of predicted and observed phenomena due to time-dependent fields 
have been reported in recent publications \cite{Scholz,Vaezi,Lop,Ding,Perf,Rocha,Ken,Sav}.  

In particular, Ref.~\cite{Scholz} argues that time-periodic spin-orbit interactions 
lead to an interesting time evolution of the spin polarization and of the optical
conductivity. Particle transport can also be induced by a time-dependent elastic
deformation field \cite{Vaezi}, or in a.c.~driven graphene nanoribbons,
where by adopting a tight-binding model, the authors of Ref.~\cite{Rocha} found
a strong dependence of transport properties on the
geometry of the ribbon edges. Furthermore, Ishikawa \cite{Ken} studied electron
transport in graphene perturbed by a time-periodic vector potential, which results
in an enhancement of interband transitions. Finally, electron transport and current resonances
in the presence of a time-dependent scalar potential barrier have been
studied in Ref.~\cite{Sav}, where a resonant enhancement of both
electron backscattering and the currents across and along the
barrier was reported when the modulation frequencies satisfy certain
resonance conditions. 

Ultrafast dynamics and particle transport in
graphene driven by ultrashort optical pulses have also been studied recently
\cite{Rusin,Liu1,Liu,Roberts,Calvo,Li,Kelardeh}. 
The experimental observation of a
bright broadband photoluminescence in graphene interacting with
femtosecond laser pulses was reported in Ref.~\cite{Liu}.
Moreover, the authors of Ref.~\cite{Kelardeh} have studied the
modification of the bandstructure under ultrashort optical
pulses and the carrier dynamics caused by the optical response of graphene,
arguing that the electron dynamics in the time-dependent electric field of the laser pulse
becomes irreversible, with a large residual conduction band population.
In addition, the formation of a laser-induced band gap was discussed in Ref.~\cite{Calvo}.

\begin{figure}
\resizebox{0.9\columnwidth}{!}{
\includegraphics{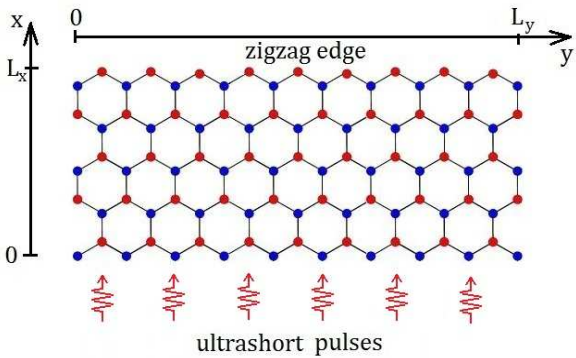} }
\caption{Sketch of a kicked zig-zag graphene nanoribbon of width $L_x$ and length $L_y$.
Ultrashort periodic pulses applied to the nanoribbon act as kicking potential. 
\label{fig1}}
\end{figure}

In this paper, we study electron transport in graphene nanoribbons
interacting with an external time-periodic scalar potential 
represented by a sequence
of $\delta$-kicks. The setup is  schematically shown in Fig.~\ref{fig1}.
 Such a potential could be created by applying laser pulses to free-standing
samples.  Using the exact solution of the time-dependent Dirac equation within 
one kicking period, we compute the transport properties of the system, such as the 
probability current  and the optical conductivity, as a function of time.
As we have discussed above, periodically driven graphene could be 
realized through the interaction with a.c.~voltages~\cite{Rocha},  pulsed laser fields \cite{Las},
surface acoustic waves \cite{Saw}, or time-periodic straining \cite{Vaezi}.
Here we focus on the case of ultrashort optical pulses \cite{Rusin,Liu1,Liu,Roberts}. 

Let us mention at this stage that some time ago,
both the classical and the quantum dynamics of systems interacting with 
a delta-kicking potential have been extensively studied
in the context of nonlinear dynamics and quantum chaos theory \cite{Casati,Sankar,Izr}.
A remarkable feature of periodically driven quantum systems is the  
quantum localization phenomenon, which implies a suppression of the 
growth of the average kinetic energy with time;
for the corresponding classical system, this energy grows linearly in time.
However, the case of delta-kicked graphene nanoribbons is more complicated due to the
kicking-field-induced  modification of the graphene bandstructure.

Such a  modification of the  bandstructure is the underlying
reason for the interesting electronic and transport phenomena found in driven graphene.
The main effect of the driving force is to cause inter- and intra-band transitions,
leading to excitation and "ionization" of valence-band electrons to the conduction band. 
Another effect caused by 
driving fields in graphene is a band-gap opening or widening \cite{Calvo},
which allows one to tune the electronic properties 
using an external time-dependent field. Below, we analyze the time evolution of
the quasi-bandstructure and of the optical conductivity
in delta-kicked graphene nanoribbons. We find that the quasienergy levels exhibit
intra-band crossings and inter-band anticrossings, where the time-dependent effective density of states reaches a local maximum when the anticrossings take place.  This may increase the number of charge carriers in the conduction band, with a subsequent increase of the current and
of the optical conductivity. Indeed, as  is shown by our analysis of 
the time-dependent conductivity in Sec.~\ref{sec3}, depending on the kicking
parameters, the conductivity may monotonically grow in time, while in other kicking regimes, such a growth is suppressed.

The remainder of this paper is organized as follows. In Sec.~\ref{sec2} we briefly
recall the Dirac equation for graphene zigzag nanoribbons, basically 
following the approach of Brey and Fertig \cite{BF}, and 
discuss the solution of the time-dependent Dirac equation in the presence of
the $\delta$-kicking potential.  This solution is then
utilized to compute the time evolution of the quasi-bandstructure, and, in
 Sec.~\ref{sec3},  the optical conductivity in different 
kicking regimes. Finally,
Sec.~\ref{sec4} contains some concluding remarks. Below we often use units where $\hbar=1$.

\section{Kicked graphene nanoribbon} \label{sec2} 

\subsection{Unperturbed Hamiltonian}\label{sec2a}

In this work we study the electronic behavior of kicked zigzag graphene 
nanoribbons, see Fig.~\ref{fig1} for an illustration, 
within the Dirac equation approach \cite{BF}. It is well established that 
low-energy quasiparticles in an extended graphene sheet are accurately described by the
 massless two-dimensional (2D) Dirac Hamiltonian \cite{review1}
\begin{equation}\label{h0}
H_0 = v_F\left(\begin{array}{cc} 0 & p_x -ip_y\\p_x+ip_y & 0
\end{array} \right),
\end{equation}  
where $p_x =-i\partial_x$, $p_y=-i\partial_y$,
and $v_F\approx 10^6$~m$/$sec denotes the Fermi velocity.
The $2\times 2$ matrix structure of $H_0$ is with respect to sublattice space, 
corresponding to the ($A/B$-type) basis atoms of graphene's honeycomb lattice \cite{review1}.  
Since we do not take into account electron-electron
interaction effects here, the two different valleys ($K$ points) as well as the 
two spin projections decouple, and we can focus on a single-valley spinless system 
in Eq.~\re{h0}.  The spinor eigenstates of the zigzag nanoribbon with periodic
boundary conditions along the longitudinal $y$-direction, see Fig.~\ref{fig1}, 
are written as 
\begin{equation} \label{WF1} 
\psi(x,y) = \frac{e^{ik_y y}}{\sqrt{L_y}} \Phi(x),\quad
\Phi(x)= \left ( \begin{array}{c} \phi_A(x)  \\
\\\phi_B(x) \ \end{array} \right), 
\end{equation}
where $k_y$ is the conserved wave number along the $y$-direction. Periodic
boundary conditions yield $k_y=2\pi n_y/L_y$ with integer $n_y$.
To take into account the zigzag edges at $x=0$ and $x=L_x$,  
where $L_x$ is the width of the nanoribbon, 
we impose the boundary conditions \cite{BF}
\begin{equation} \label{bc1} 
\phi_A(L_x) = \phi_B(0) = 0, 
\end{equation}
and put $0\le x\le L_x$ henceforth.  

Next we summarize the spinor solutions $\psi_n(x,y)$ solving
the stationary Dirac equation for eigenenergy $E_n$ in the absence of
the kicking potential,
\begin{equation} \label{Dir2}
H_0\psi_n(x,y) =E_n\psi_n(x,y),  \quad n=(n_x,n_y),
\end{equation}
where the integer $n_x$ serves as a band index and $n_y$ parametrizes $k_y$.
The boundary conditions \re{bc1} imply that the eigenvalues of Eq.~\re{Dir2} 
are obtained from  the transcendental equation \cite{BF}
\begin{equation} \label{spect}
\frac{k_y-z}{k_y+z}=e^{-2L_xz}, 
\end{equation}
which admits two type of solutions, namely (i) confined
modes (standing waves), and (ii) surface states. 

We start by discussing solutions of type (i), which are
 purely imaginary, $z_{n_x}=ik_x$, and lead to the eigenenergies 
\begin{equation}\label{ene}
E_n=\pm v_F\sqrt{k_x^2+ k_y^2 },
\end{equation} 
where the upper (lower) sign corresponds to the conduction (valence) band.
Equation \re{spect} now simplifies to 
\begin{equation} \label{eig}
k_y = \frac{k_x}{\tan(k_x L_x)} , 
\end{equation}
and solutions for $k_x$ (labeled by $n_x=1,2,\ldots$)
correspond to confined modes. The respective  
eigenstate \re{WF1}, with  the energy $E_n$ in Eq.~\re{ene},
has the transverse wavefunction
\begin{equation} \label{wf2} 
\Phi_n(x)  = {\cal N}_n \left( \begin{array}{c} \sin(k_x x)  
\\ \pm \frac{i}{E_n} [- k_x \cos(k_x x) + k_y \sin(k_x x)]
\end{array} \right),
\end{equation} 
with normalization constant ${\cal N}_n$.

\begin{figure}
\resizebox{0.9\columnwidth}{!}{\includegraphics{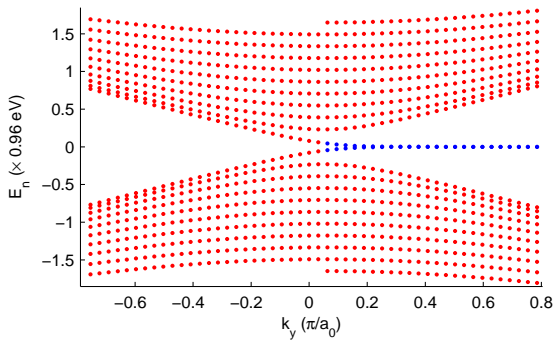} }
\caption{Band structure of unperturbed graphene nanoribbon. 
The red dots indicate eigenenergies of confined modes, and the blue dots
correspond to surface states.  The nanoribbon has width 
$L_x =4.92$~nm and length $L_y =12.3$~nm, and $a_0=0.246$~nm
is the lattice unit.
 \label{fig2}}
\end{figure}

Next we turn to surface state [type (ii)] solutions.
These correspond to purely real $z=k_x$ in Eq.~\re{spect}, 
where the eigenstate reads
\begin{equation} \label{wf4} 
\Phi_n(x) = {\cal N}_n^\prime \left( \begin{array}{c} 2\sinh(k_x x)  \\
\frac{1}{E_n}\left[k_y \sinh(k_x x)-k_x\cosh(k_x x)\right] \end{array} \right).
\end{equation}
The eigenenergy is now given by
\begin{equation}
E_n=\pm \sqrt{k_y^2-k_x^2},
\end{equation}
and ${\cal N}^\prime_n$ is another normalization constant.
The surface state energies equal zero for sufficiently large positive $k_y$, 
but they are absent for $k_y<0$.  In Fig.~\ref{fig2}, the resulting 
bandstructure of a typical graphene zigzag nanoribbon is plotted. 

\subsection{Including the kicking potential}\label{sec2b}

We are now ready to include the external driving potential.  We consider
a periodic sequence of delta-kicks of 
kicking strength $\varepsilon$ and period $T$.  (No confusion with the symbol for temperature
should arise here; we always consider the zero-temperature limit.) 
Writing $H=H_0+ V {\rm diag}(1,1)$, the additional term is given by the time-periodic scalar
potential, 
\begin{equation}\label{kick}
V(x,t) = \varepsilon \cos \left(2\pi x/\lambda\right) \sum_{l=0}^{\infty} \delta(t-lT),
\end{equation}
where $\lambda$ is the wavelength of the kicking pulse. 
Experimentally, such delta-kicks could be realized
by standing-wave laser pulses \cite{Raizen,Ullah1,Ullah2}, or by half-cycle 
laser pulses \cite{Matos}.
For instance, the delta-kicked quantum rotor, representing a well-known paradigm of 
quantum chaos theory, can be experimentally realized in ultracold atoms that interact with the
periodic standing wave of a near-resonant laser field \cite{Raizen}.
Significant progress concerning the experimental realization of graphene 
interacting with ultrashort laser pulses has also been reported recently
 \cite{Liu1,Liu,Roberts}. 
Combining the experimental methods in Refs.~\cite{Raizen,Ullah1,Ullah2,Matos} 
with those in Refs.~\cite{Liu1,Liu,Roberts}
could allow to implement the delta-kicked graphene nanoribbon discussed here in the lab.

The dynamics of a state $\Psi=\Psi(x,y,t)$ is then 
governed by the time-dependent 2D 
Dirac equation, $i \partial_t \Psi = H\Psi$. 
To solve this equation, we expand $\Psi(x,y,t)$ in terms of the
complete set of eigenfunctions of the unperturbed graphene
zigzag nanoribbon discussed in Sec.~\ref{sec2a},
\begin{equation} \label{WF}
\Psi(x,y,t) =\sum_{n} A_{n}(t) \psi_{n}(x,y), 
\end{equation} 
where $n=(n_x,n_y)$  and
the summation implicitly  includes the $\pm$ 
sign for the conduction and valence band, respectively. 
To ensure normalization, the initial values (at time $t=0$) of the 
complex-valued expansion coefficients $A_n(t)$ in Eq.~\re{WF} satisfy the condition
\begin{equation} \label{norm}
\sum_n |A_{n}(0)|^2 = 1.
\end{equation}
Within one time period, the amplitude $A_n$ then follows the time evolution
\begin{equation} \label{evol} 
A_{n}(t+T) = \sum_{n'} V_{nn'} e^{-i E_{n'} T} A_{n'}(t) 
\end{equation} 
where $E_n$ is the unperturbed eigenenergy of the respective mode,
 see Sec.~\ref{sec2a},
and we define the matrix elements 
\begin{equation} 
V_{nn'} = \int_0^{L_x} dx \int_0^{L_y} dy\ \psi^\dagger_{n'}(x,y) 
e^{i\varepsilon \cos (2\pi x/\lambda)} \psi_{n} (x,y) ,
\end{equation} 
where nonzero matrix elements exist only for $n_y=n_y'$ due 
to the translation invariance in $y$-direction.
In calculating these matrix elements, we use a well-known Bessel function
expansion formula for the exponential term.

In numerical calculations, one may choose only a few non-zero initial coefficients $A_n(0)$ 
subject to Eq.~\re{norm}. In particular, we tested the impact of using different
choices for $A_{n}(0)$, such as randomly chosen or equally distributed
values.  All choices were found to give qualitatively similar results for the time-evolved state
$\Psi(t)$ after many kicks.  For the calculations presented below, 
we chose a random distribution for the coefficients $A_n(0)$ containing $\approx 15$
non-zero entries, where we take into account only states with energy $E_n$
below the Fermi level. The Fermi level is here assumed at the neutrality point, i.e.,
$E_F=0$, in order to maximally emphasize the Dirac fermion nature of the graphene nanoribbon.  
Our procedure for choosing the initial values for the $A_n$ coefficients 
mimics the zero-temperature average over the filled Fermi sea.
We have carefully checked that different initial values lead to the same physical results
after a short initial transient.

\begin{figure}[t!]
\resizebox{1\columnwidth}{!}{%
\includegraphics{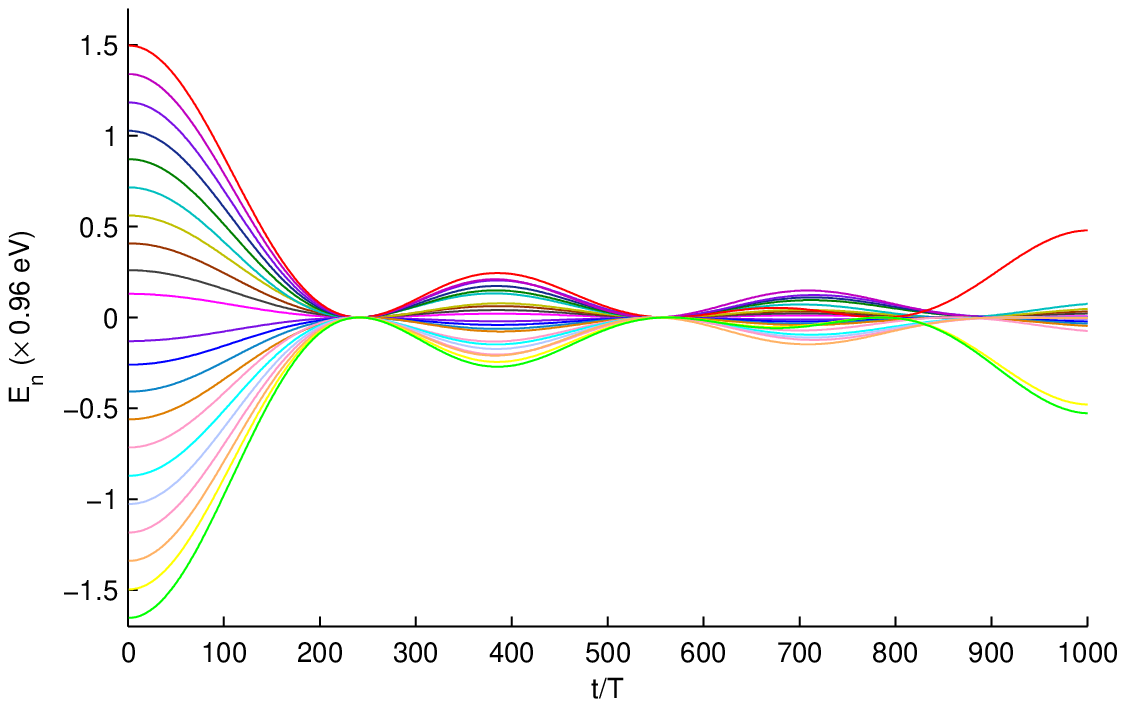}}
\resizebox{1\columnwidth}{!}{%
\includegraphics{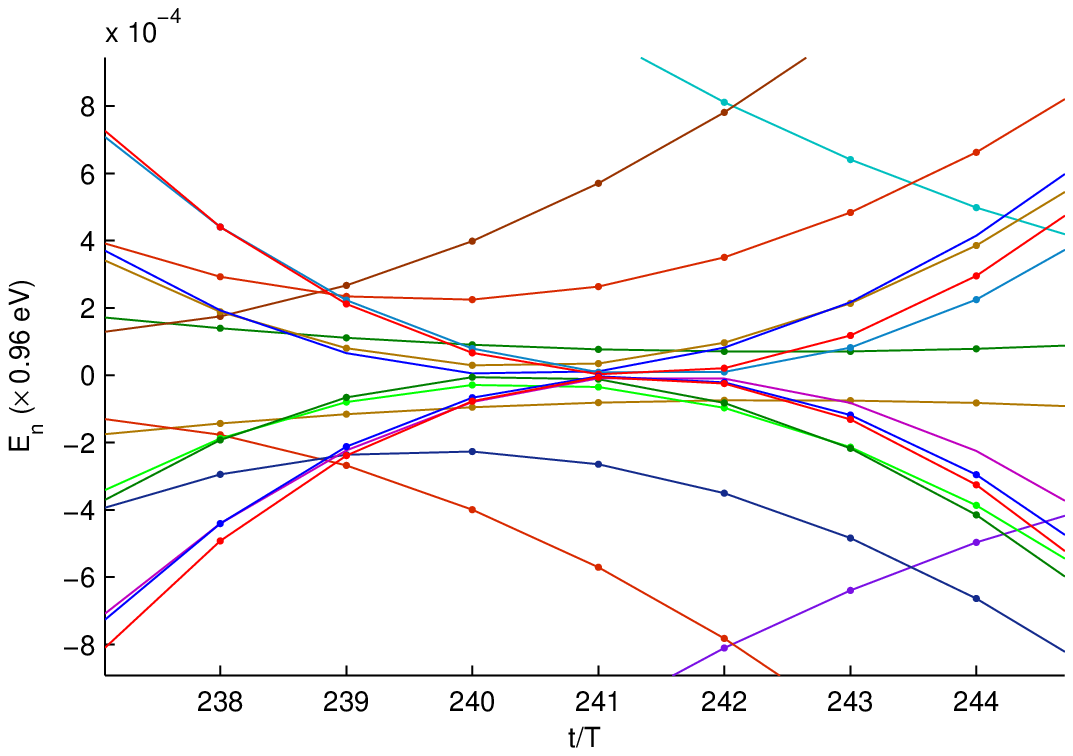}}
\caption{\label{fig3} The upper panel shows the time evolution of a few selected
quasienergy levels.  The lower panel shows the same but on a magnified scale
near the first level (anti-)crossing.  The ribbon width $L_x$ and
length $L_y$ were chosen as in Fig.~\ref{fig2}.  
The kicking strength is $\varepsilon =2.7\times 10^{-13}$~eVs, with
period $T=0.1$~fs and wavelength $\lambda=2.46$~nm.}
\end{figure}

Given the wave function, one can compute different
characteristics of the carrier dynamics and, in particular, 
investigate charge transport in a kicked graphene nanoribbon. 
In Fig.~\ref{fig3}, the time evolution of quasienergy levels is shown for a
few selected states. 
 We observe several crossings of the levels within the conduction (or
within the valence) band.  However, levels coming from different bands
exhibit anticrossing, where levels closely approach each other up
to some time when they start to separate again. After a 
certain number of kicks, one can then again observe crossings or 
anticrossings, where intra- and interband transitions become more frequent.  
Since initially the valence band is filled, this 
can lead to an increase in the number of electrons in the conduction band, and
thereby to current flow. A related enhancement of intra- and inter-band
transitions has also been reported in Ref.~\cite{Kelardeh} 
for graphene subject to ultrashort laser pulses. When
the quasienergy levels separate  from each other again after the crossing or 
anticrossing, intra- and interband transitions become less frequent, and one
can expect a decrease in the current.
Such features indeed appear in the conductivity, as we study next.

\section{Optical conductivity}
\label{sec3}

The interaction of external electromagnetic fields with solids generally causes a
modification of their electronic properties and, in particular, of 
the bandstructure \cite{Kelardeh,Tzoar}. Using the solution of the
time-dependent Dirac equation for delta-kicking potential discussed in Sec.~\ref{sec2},
one can compute such modifications in the bandstructure, see Fig.~\ref{fig3}.
In this section, we focus on the optical conductivity of our system, 
which represents an important observable of experimental interest
and can provide precious insights about the transport
mechanisms at play in kicked graphene nanoribbons.

Within linear response theory, the Kubo formula yields for the 
diagonal elements of the time-dependent conductivity tensor ($\alpha=x,y$)
\cite{altland}
\begin{eqnarray}\label{con1}
\sigma_{\alpha\alpha} (x,y;t,\omega) & =  &
\frac{e^2}{\omega} \int_{0}^{\infty}
d\tau e^{-i\omega \tau} \\ \nonumber
&\times& \left\langle [ J_\alpha(x,y,t), J_{\alpha}(x,y,t -\tau)]\right\rangle,
\end{eqnarray}
where $[,]$ denotes the commutator and 
the particle current density along the $\alpha$-direction is \cite{review1}
\begin{equation}\label{cur1}
J_\alpha(x,y,t) = v_F\Psi^\dagger(x,y,t)\sigma_\alpha \Psi(x,y,t),
\end{equation}
with standard Pauli matrices $\sigma_{\alpha=x,y}$ acting in sublattice space.
The average in Eq.~\re{con1} is taken with respect to the filled Fermi sea at the 
initial time $t=0$, present before the kicking potential is switched on.
In Eq.~\re{con1}, we focus on the long-wavelength limit 
by probing the two current operators appearing in the Kubo formula 
at the same point in space.
In terms of the expansion coefficients $A_{n}(t)$ appearing in the 
expansion \re{WF}, Eq.~\re{cur1} takes the form
\begin{equation}\label{cur2}
J_\alpha(x,y,t)=v_F \sum_{nn'} 
A_{n'}^*(t) A_n(t) \psi_{n'}^\dagger(x,y) \sigma_\alpha 
\psi_n^{}(x,y),
\end{equation}
where $A^*$ denotes the complex conjugate of $A$.
Inserting Eq.~\re{cur2} into Eq.~\re{con1}, the conductivity at time $t$
follows as a lengthy (but straightforwardly obtained) expression 
involving the time-dependent coefficients $\{ A_n(t') \}$ for $0<t'<t$. 
As described in Sec.~\ref{sec2b}, the zero-temperature average over
 the filled Fermi sea 
is implemented by choosing suitable initial values for those coefficients.
Below we discuss the time-dependence of the conductivity 
averaged over the sample area and taken in the $\omega\to 0$ limit,
\begin{equation} \label{sigmafin}
\sigma_{\alpha\alpha} (t) = \lim_{\omega\to 0} {\rm Re}
\int_{0}^{L_y} \frac{dy}{L_y} \int_0^{L_x} \frac{dx}{L_x}  
\sigma_{\alpha\alpha} (x,y; t,\omega).
\end{equation}

\begin{figure}[t!]
\resizebox{1\columnwidth}{!}{%
\includegraphics{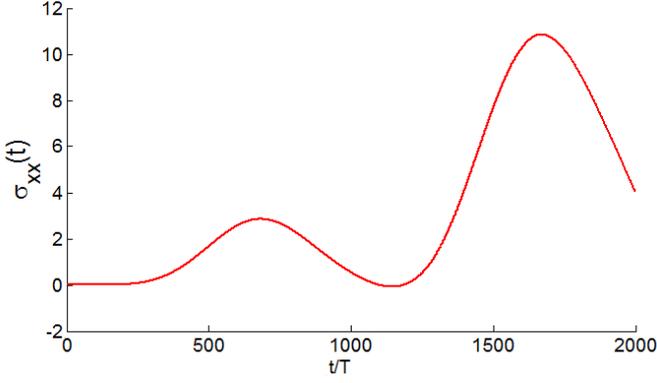}}
\vspace{3mm} \caption{Conductivity $\sigma_{xx}(t)$ in units of $e^2/\hbar$,
 see Eq.~\re{sigmafin}, as a
function of time for a kicked graphene nanoribbon.  Parameters
are as in Fig.~\ref{fig3} but with kicking time period $T=1$~fs.   \label{fig4}}
\end{figure}

Figure~\ref{fig4} shows numerical results for $\sigma_{xx}(t)$
for representative kicking potential parameters.  
We observe that $\sigma_{xx}(t)$ grows during
some initial time interval, followed by a suppression of the growth 
along with the development of oscillatory behavior.
These features can be linked to the presence of crossings and
anticrossings in the bandstructure, as discussed above.

\begin{figure}[t!]
\resizebox{1\columnwidth}{!}{
\includegraphics{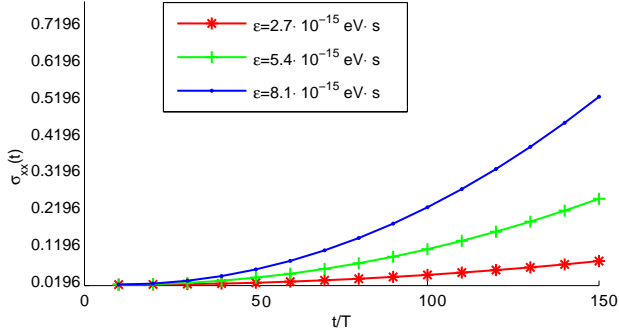}}
\caption{ Same as Fig.~\ref{fig4} but for several values of the 
kicking strength $\varepsilon$,
and showing only the initial time period ($t<150 T$).  
\label{fig5}}
\end{figure}

Next, Fig.~\ref{fig5} presents $\sigma_{xx}(t)$ for a fixed 
kicking period, $T=1$~fs, but now for
different values of the kicking strength $\varepsilon$.
For the shown regime of rather short times, the conductivity 
monotonically grows  in time. Clearly, for  larger kicking strength,
$\sigma_{xx}(t)$ grows more rapidly, where the slope of the
growth is approximately proportional to  $\varepsilon$.
However, as seen in Fig.~\ref{fig4}, after a longer time span,
the conductivity will be suppressed again.  Nonetheless, 
the variation of the kicking strength allows for a 
considerable tunability of the conductivity.

\begin{figure}[t!]
\resizebox{1\columnwidth}{!}{%
\includegraphics{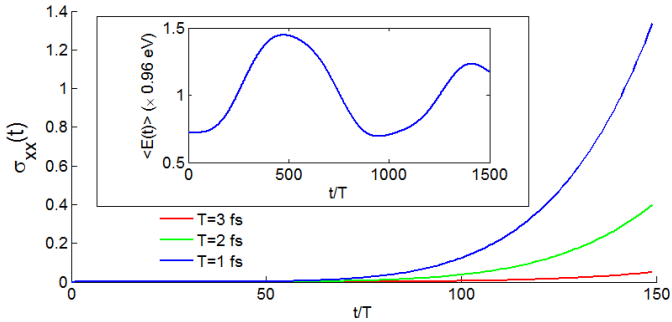}}
\caption{Same as Fig.~\ref{fig5} but for different kicking periods $T$
and the fixed kicking strength $\varepsilon=2.7\times 10^{-13}$~eVs.
The inset shows the time dependence of the average kinetic energy 
on a longer time scale for $T=1$~fs. \label{fig6} }
\end{figure}

The dependence of the conductivity on a variation of the kicking period
$T$ (with fixed strength $\varepsilon)$ is illustrated by
Fig.~\ref{fig6}.  We find that $\sigma_{xx}(t)$ monotonically grows within
the considered time interval, although at larger times (not shown)
the conductivity decreases and becomes quasi-oscillatory again,
as displayed before in Fig.~\ref{fig4}.
The inset of Fig.~\ref{fig6} also shows the time-dependence of the average
kinetic energy,
\begin{eqnarray}\nonumber
\langle E(t)\rangle &=& \int_0^{L_x}dx  \int_0^{L_y} dy \
\Psi^*(x,y,t) H_0 \Psi(x,y,t)  \\
&=& \sum_n |A_n(t)|^2 E_n.
\end{eqnarray}
This quantity exhibits time-periodic behavior, again reflecting
the periodic appearance of band crossings as illustrated by Fig.~\ref{fig3}.
Such a behavior is different from a localization-induced
saturation expected from quantum chaos theory but also
differs from the simple monotonic growth expected on classical
grounds \cite{Casati,Sankar,Izr}. 

\begin{figure}[t!]
\resizebox{1\columnwidth}{!}{%
\includegraphics{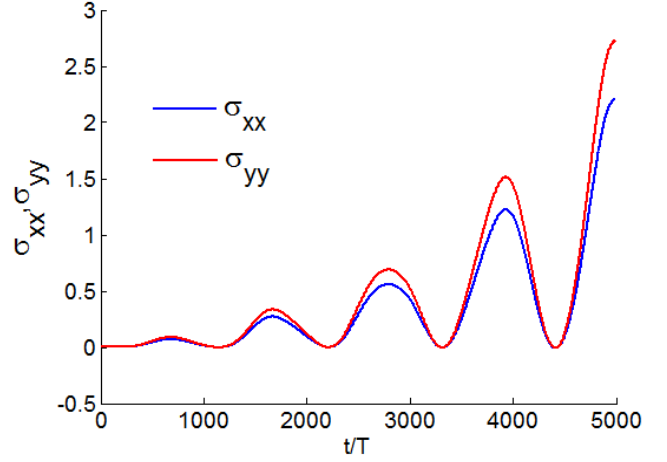}}
\caption{ Time-dependence of the longitudinal conductivities $\sigma_{xx}(t)$ 
and $\sigma_{yy}(t)$ (in units of $e^2/\hbar$) 
along the $x$- and $y$-direction, respectively,
for the parameters in Fig.~\ref{fig4}. 
\label{fig7} }
\end{figure}

Next, in Fig.~\ref{fig7}, we compare the different components of the 
conductivity, namely $\sigma_{xx}(t)$ and $\sigma_{yy}(t)$. 
Although the kicking force acts along $x$-direction, we find 
very similar values for the conductivities in both directions. 
The spectral rearrangement caused by the kicking force is thus 
quite efficient in also inducing current fluctuations 
in the transverse ($y$) direction.

\begin{figure}[t!]
\resizebox{1\columnwidth}{!}{%
\includegraphics{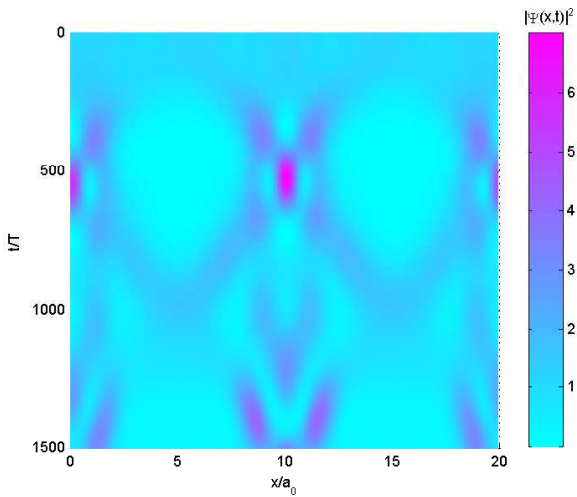}}
\caption{Probability density as a function of time
and coordinate for the parameters in Fig.~\ref{fig4} (note
that $L_=20a_0$).
 \label{fig8}}
\end{figure}

Finally, Fig.~\ref{fig8} presents the spatio-temporal evolution of the
probability density, $|\Psi(x,y,t)|^2$.  Since the result is homogeneous
along the $y$-direction, we show it as 2D color-scale plot in the $(x,t)$
plane; note the symmetry under spatial 
reflection with respect to the ribbon midpoint $x=L_x/2$.
Figure \ref{fig8} provides additional 
information about the possibility of spatio-temporal
quantum localization of electrons in the kicked graphene nanoribbon.  
In fact, we conclude from Fig.~\ref{fig8} that the carriers are not 
fully localized inside the ribbon, although signatures of 
localization near the midpoint are visible for finite time spans.

\section{Concluding remarks}
\label{sec4}

In this paper, we have studied time-dependent particle transport in
graphene zigzag nanoribbons driven by an external 
time-periodic $\delta$-kicking potential.  The time-dependent 
Dirac equation can be solved exactly within a single kicking period, 
and numerical iteration of this solution provides access to the
wave function at arbitrary time.  Using this wave function, we have computed
the time-dependent optical conductivity (and other quantities). 
The conductivity is observed to initially grow as a function of time 
up to certain time, after which the conductivity decreases and ultimately shows
quasi-oscillatory behavior. 

We find it rather remarkable that  
by judiciously choosing the strength $\varepsilon$ and 
the time period $T$ of the kicking field, one can achieve almost
arbitrary results for the 
oscillation period and for the amplitude of the conductivity.
In particular, it is possible to choose parameters such that
the initial increase extends for very long time.
The described behavior of $\sigma_{\alpha\alpha}(t)$ can be linked to the existence of (anti-)crossings in the quasienergy bands of the driven system.  

The model  studied in our work could be  realized in zigzag ribbons
made of monolayer graphene samples that are exposed to 
standing-wave ultrashort laser pulses, such as those discussed
 in Refs.~\cite{Raizen,Ullah1,Ullah2}.
The above results also may help in solving the problem of tunable charge
transport in graphene-based electronic devices. 

\section{Acknowledgments}
This  work has  been supported by the Volkswagen-Stiftung.


\begin{thebibliography}{150}
\bibitem{review1} A.H. Casto Neto, F. Guinea, N.M.R. Perez, 
K.S. Novoselov, and A.K. Geim, Rev. Mod. Phys.  {\bf 81}, 109 (2009).
\bibitem{review2} N.M.R. Peres, Rev. Mod. Phys.  {\bf 82}, 2673 (2010).
\bibitem{review3} S. Das Sarma, S. Adam, E. H. Hwang, and E. Rossi, 
Rev. Mod. Phys.  {\bf 83}, 407 (2011).
\bibitem{review4} A.V. Rozhkov, G. Giavaras, Y.P. Bliokh, 
V. Freilikher, and F. Nori, Phys. Rep. {\bf 503}, 77 (2011).
\bibitem{review5} A.H. Castro Neto and K. Novoselov, Rep. Prog. Phys. {\bf 74}, 082501 (2011).
\bibitem{review6} 
V.N. Kotov, B. Uchoa, V.M. Pereira, F. Guinea, and A.H. Castro Neto, 
Rev. Mod. Phys.  {\bf 84}, 1067 (2012).
\bibitem{review7} J. G\"{u}ttinger, F. Molitor, C. Stampfer, S. Schnez, 
A. Jacobsen, S. Dr\"{o}scher, T. Ihn, and K. Ensslin, 
Rep. Prog. Phys.  {\bf 75}, 126502 (2012).
\bibitem{Scholz} A. Scholz, A. Lopez, and J. Schliemann, Phys. Rev. B  {\bf 88}, 045118 (2013).
\bibitem{Vaezi} A. Vaezi, N. Abedpour, R. Asgari, A. Cortijo, and M.A.H. Vozmediano, 
Phys. Rev. B  {\bf 88}, 125406 (2013).
\bibitem{Lop} A. Lopez, Z.Z. Sun, and J. Schliemann, Phys. Rev. B  {\bf 85}, 205428 (2011).
\bibitem{Ding} K.-H. Ding, Z.-G. Zhu, and J. Berakdar, Phys. Rev. B  {\bf 84}, 115433 (2011).
\bibitem{Perf} E. Perfetto, G. Stefanucci, and M. Cini, Phys. Rev. B  {\bf 82}, 035446 (2010).
\bibitem{Rocha} C.G. Rocha, L.E.F. Foa Torres, and G. Cuniberti, Phys. Rev. B  {\bf 81}, 115435 
(2010).
\bibitem{Ken} K.L. Ishikawa, Phys. Rev. B  {\bf 82}, 201402, (2010).
\bibitem{Sav} S.E. Savelev, W.~H\"ausler, and P. H\"anggi, Phys. Rev. Lett.  
{\bf 109}, 226602 (2012).
\bibitem{Rusin} T.M.Rusin and W. Zawadzki, Phys. Rev. B  {\bf 80}, 045416 (2009).
\bibitem{Liu1} C.H. Lui, K.F. Mak, J. Shan, and T.F. Heinz, Phys. Rev. Lett. {\bf 105}, 127404
(2010).
\bibitem{Liu} W.-T.Liu, S. W. Wu, P. J. Schuck, M. Salmeron, Y. R. Shen, and F. Wang, 
Phys. Rev. B  {\bf 82}, 081408 (2010).
\bibitem{Roberts} A. Roberts, D. Cormode, C. Reynolds, T. Newhouse-Illige, B.J. LeRoy, and 
A.S. Sandhu, Appl. Phys. Lett. {\bf 99}, 051912 (2011).
\bibitem{Calvo} H.L. Calvo,  H.M. Pastawski, S. Roche, and L.E.F. Foa Torres, 
Appl. Phys. Lett.  {\bf 98}, 232103 (2011).
\bibitem{Li} T. Li, L. Luo, M. Hupalo, J. Zhang, M.C. Tringides, J. Schmalian, and 
J. Wang, Phys. Rev. Lett. {\bf 108}, 167401 (2012).
\bibitem{Kelardeh} H.K. Kelardeh, V. Apalkov, and M.I. Stockman,  preprint arXiv:1401.5786.
\bibitem{Las} H.K. Avetissian, A.K. Avetissian, G.F. Mkrtchian, and Kh.V.  Sedrakian, Phys. Rev. B  {\bf 85}, 115443 (2012).
\bibitem{Saw} P. Thalmeier, B. Dora, and K. Ziegler, Phys. Rev. B  {\bf 81}, 041409(R) (2010).
\bibitem{Casati} G. Casati {\sl et al.}, in \textit{Lecture Notes in Physics} {\bf 93}, 334 
(Springer, Berlin, 1979).
\bibitem{Sankar} R. Sankaranarayanan and V.B. Sheorey, Phys. Lett. A  {\bf 338}, 288 (2005).
\bibitem{Izr} G.M. Izrailev, Phys. Rep. {\bf 196}, 299 (1990).
\bibitem{BF} L. Brey and H. A. Fertig, Phys. Rev. B  {\bf 73}, 235411 (2006).
\bibitem{Raizen} F.L. Moore, J.C. Robinson, C.F. Bharucha, B. Sundaram, and M.G. 
Raizen, Phys. Rev. Lett. {\bf 75}, 4598 (1995).
\bibitem{Ullah1} A. Ullah and M.D. Hoogerland, Phys. Rev. B {\bf 83}, 046218 (2012).
\bibitem{Ullah2} A. Ullah, S.K. Ruddell, J.A. Currivan, and M.D. Hoogerland, 
Eur. Phys. J. D {\bf 66}, 315 (2012).
\bibitem{Matos} A. Matos-Abiague and J. Berakdar, Phys. Rev. B {\bf 69}, 155304 (2004).
\bibitem{Tzoar} N. Tzoar and J.I. Gersten, Phys. Rev. B  {\bf 12}, 1132 (1975).
\bibitem{altland} A. Altland and B.D. Simons, 
\textit{Condensed Matter Field Theory}, 2nd. edition
(Cambridge University Press, Cambridge UK, 2010).
\end{thebibliography}
\end{document}